\documentclass{article}
\usepackage{spconf,amsmath,graphicx,amsfonts,mathtools}

\title{Music Enhancement with Deep Filters: \\ A Technical Report for The ICASSP 2024 Cadenza Challenge}
\name{Keren Shao \qquad  Ke Chen \qquad  Shlomo Dubnov
\thanks{We would like to thank IRCAM — Project REACH for supporting this project.}}
\address{University of California, San Diego, USA}
\begin{document}
\ninept
\maketitle
\begin{abstract}
In this challenge, we disentangle the deep filters from the original DeepfilterNet and incorporate them into our Spec-UNet-based network to further improve a hybrid Demucs (hdemucs) based remixing pipeline. The motivation behind the use of the deep filter component lies at its potential in better handling temporal fine structures. We demonstrate an incremental improvement in both the Signal-to-Distortion Ratio (SDR) and the Hearing Aid Audio Quality Index (HAAQI) metrics when comparing the performance of hdemucs against different versions of our model.
    
\end{abstract}

\section{Introduction}
\label{sec:intro}

The ICASSP 2024 Cadenza Challenge \cite{ICASSP2024-Cadenza} focuses on improving music perception for individuals with hearing aids. While general audio enhancement and speech enhancement are well-researched topics and are advancing rapidly with the rise of deep learning techniques \cite{yu2022high, hao2021fullsubnet}, enhancement techniques specifically targeting people with hearing disabilities have received less attention.

As shown in Figure 1, in this challenge, the enhancement pipeline separates the mixtures at the hearing aid's microphones (mixtures-at-mic) into drums, bass, other and vocal stems using pre-trained baseline systems such as hdemucs \cite{defossez2021hybrid}, and outputs an enhanced remix given a set of gains $\{G_{\text{drums}}, G_{\text{bass}}, G_{\text{other}}, G_{\text{vocal}}\}$ and the listener's characteristic (NALR). Note that per challenge rules, our design does not have access to the preprocessing stages before the mixture-at-mic.

The remix is then compared against the ground-truth remix. This ground-truth remix is generated by processing individual ground truth stems from the MUSDB18 dataset \cite{musdb18-hq} in the same manner as the predicted stems. The evaluation metric used in this challenge is HAAQI \cite{kates2015hearing}. Unlike conventional metrics such as SDR, HAAQI focuses more on the envelope and temporal fine structure. It has been shown to better predict perceptual quality from the perspective of hearing aid listeners \cite{kates2015hearing}.

To further enhance this pipeline, we adopt Spec-UNet \cite{jansson2017singing} as a fine-tuning network. This allows us to leverage the solid separation performance of the baseline hdemucs.

Additionally, rather than outputting a complex ratio mask (cRM) at the end of the fine-tuning network, we borrow the concept of deep filters from \cite{schroter2022deepfilternet-main, schroter2022deepfilternet} as they take temporal information into account by filtering the spectrograms with a fixed-length window along the time axis. This approach turns out to better align with HAAQI, whose computation stresses temporal fine structures.

\begin{figure}[t]
    \centering
    \includegraphics[width=\columnwidth]{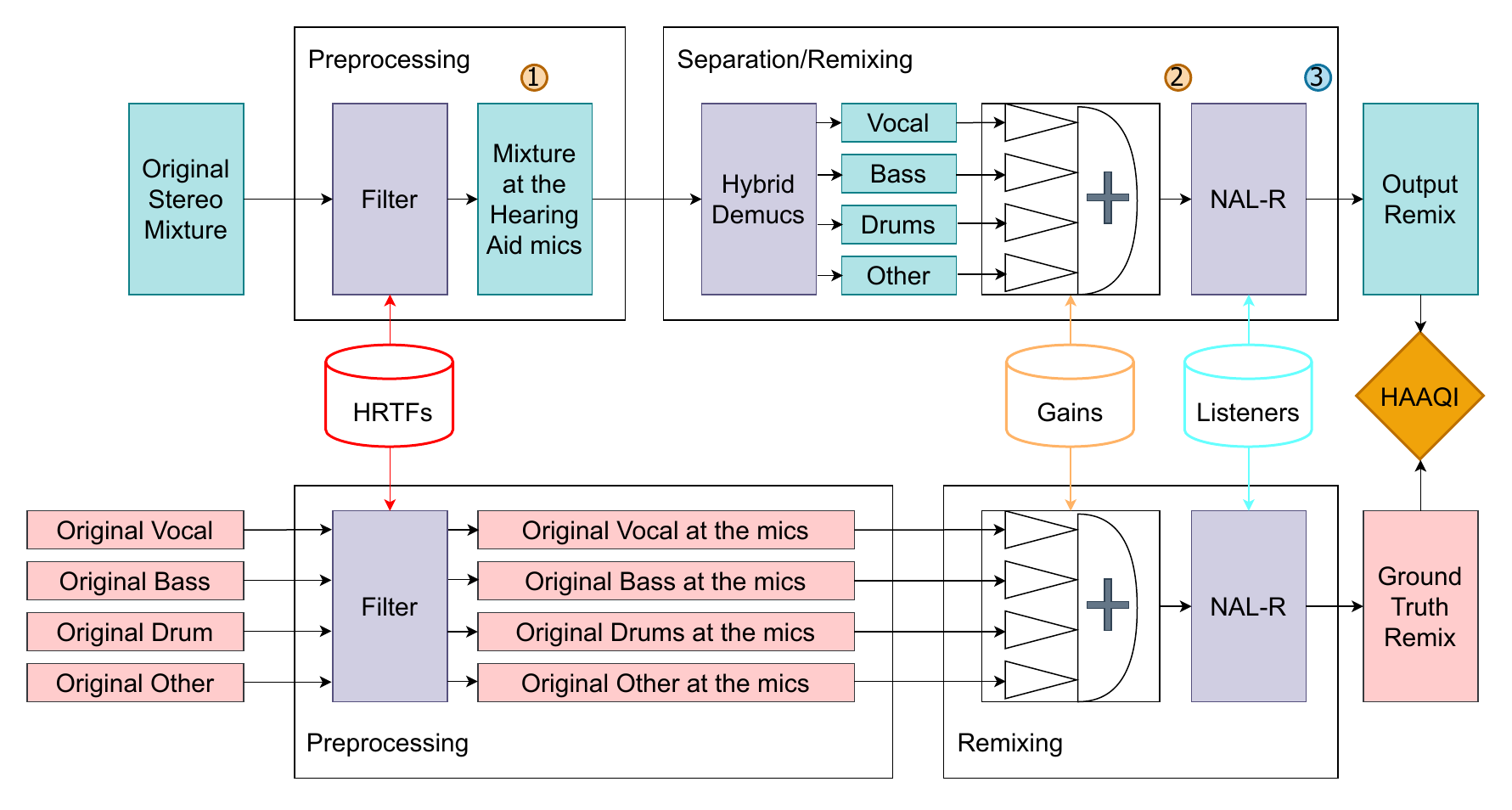}
    \vspace{-0.8cm}
    \caption{Cadenza Challenge Pipeline. The circled numbers 1, 2, and 3 indicate the components of the input to our designed model. The model output then replaces the role of component 3 and is compared against the ground truth remix.}
    \label{fig:z-cfp}
    \vspace{-0.5cm}
\end{figure}

\section{Theory}
\label{sec:pagestyle}

\begin{figure}[t]
    \centering
    \includegraphics[width=\columnwidth]{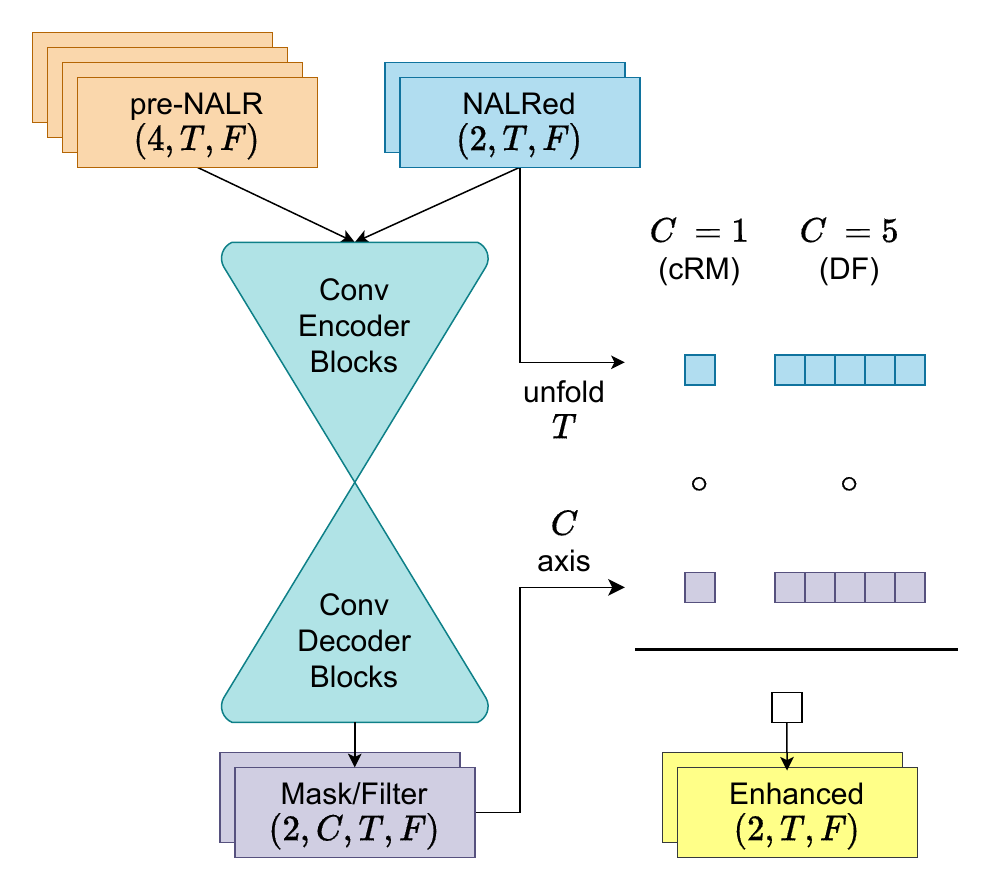}
    \vspace{-0.8cm}
    \caption{Our Model Architecture. The dot represents vector inner product. 'pre-NALR' corresponds to the stereo remix in locations 1 and 2, while 'NALRed' corresponds to the remix in location 3 in Figure 1.}
    \label{fig:z-cfp}
    \vspace{-1.04cm}
\end{figure}

\subsection{Model Input, Architecture and Loss}

As depicted in Figure 1 and Figure 2, our model takes a concatenation of the STFT of mixture-at-mic, gain-adjusted remix, and NALRed remix as input, which is of shape $\mathbb{C}^{B \times C_m \times T \times F}$, where $B$ represents the batch size, $C_m = 6$ the audio channels from pre-NALR and NALRed inputs, $T$ the time axis, and $F$ the frequency axis.

The data then flows through the Spec-UNet. The final layer produces a data shape of $\mathbb{C}^{B \times 32 \times T \times F}$, at which point we have the option to further reduce the channel count to 1 to generate a complex ratio mask or to some value $N$ to utilize it as a deep filter of order $N$, as explained in the following section. The final output shape is $\mathbb{C}^{B \times 2 \times T \times F}$.

Finally, we perform inverse STFT, and its output is trained against the ground-truth remix using the mean absolute error loss.

\subsection{Deep Filter}
Formally, given a complex ratio mask (cRM) produced by the network, we perform the following operation:
\begin{align}
O[t, f] = I[t, f] * M[t, f] \text{ for } \forall t, f
\end{align}
Here, disregarding the stereo channel dimension, $O \in \mathbb{C}^{T \times F}$ represents the final output, $I \in \mathbb{C}^{T \times F}$ represents the NALRed part of the input, and $M \in \mathbb{C}^{T \times F}$ is our mask.

For the deep filter, as explained in \cite{schroter2022deepfilternet}, we first unfold the input $I$ along the time axis with a kernel length of $N$, obtaining $I' \in \mathbb{C}^{T \times F \times N}$. Then, we let the network produce a tensor $M' \in \mathbb{C}^{N \times T \times F}$, and our output can be computed as follows:

\begin{align}
O[t, f] = I'[t, f, :] \cdot M'[:, t, f] \text{ for } \forall t, f
\end{align}
Here, the dot represents a vector dot product.



\section{Experiments}

\subsection{Datasets and Experimental setup}
We utilize the MUSDB18 dataset for the training and the evaluation of our model. The training set contains 100 music tracks with four stems (vocals, drums, bass and other). And the test set contains 50 music tracks. No data augmentation is performed, and additional datasets are not utilized. For the hyperparameters of Spec-UNet, we use the window size and the FFT size of 2048 with the hop size of 441. Since all of our inputs share a sampling rate of 44100 Hz, our output remains consistent with that. Regarding the deep filter, we set the unfolding kernel length to $N = 5$. Attempting higher kernel lengths and experimenting with two-dimensional filters produced similar performance.

In terms of general training hyperparameters, we use a batch size of 36. During training, we randomly extract 2-second clips from the pool of songs. The learning rate is set to 1e-3. All methods are implemented in PyTorch and trained on an NVIDIA A6000 GPU, with a maximum epoch of 100. Our model weighs approximately 138M. With mostly convolutional layers, a single pass of training completes in less than a day.

\subsection{Results and Discussion}

We present the results in Table 1. In the second row, we use DeepFilterNet as the enhancement network and retrain it without further adjusting its ERB (equivalent rectangular bandwidth) front end, potentially contributing to its suboptimal performance. In the last row, we only incorporate the deep filter operations from the DeepFilterNet into our model, as illustrated in Figure 2. As observed, both the SDR and HAAQI metrics exhibit incremental improvements after applying our model and replacing cRM with DF.


\begin{table}[t]
\small
\center
\renewcommand\arraystretch{1.3} 
\begin{tabular}{l|c|c}
& mean SDR (dB) & HAAQI \\
\hline
hdemucs (baseline) & 8.297 & 0.5697 \\
hdemucs + DeepFilterNet \cite{schroter2022deepfilternet} & 8.124 & \textendash \\
hdemucs + Spec-UNet (cRM) & 8.315 $\pm$ 0.013 & \textendash \\
hdemucs + Spec-UNet (DF) & 8.326 $\pm$ 0.009 & 0.5704 \\
\hline \hline
\end{tabular}

\caption{Evaluation Results on the MUSDB18 Test Set. To ensure the statistical significance of our results, we repeat the third and fourth experiments three times each and report the average.
}
\label{tab:aba_test}
\end{table}

\section{Conclusion}
In conclusion, we have designed a system that builds upon hybrid Demucs to further enhance its remixing performance. The results of SDR and HAAQI metrics not only confirm the positive effect of our design on improving perception but also indicate the challenges of generalizing beyond known listeners. We plan to investigate this issue in future work.

\bibliographystyle{IEEEbib}
\bibliography{refs}
\end{document}